\documentclass[sigchi-a, nonacm]{acmart}
\usepackage{booktabs} 
\usepackage{ccicons}  

\setcopyright{none}

\begin{document}
\title{Designing with Data: A Case Study}

\author{Teresa Castle-Green \\Stuart Reeves\\Joel E. Fischer\\Boriana Koleva}
\affiliation{%
  \institution{Mixed Reality Laboratory\\University of Nottingham}
  \city{Nottingham}
  \country{UK} }
\email{first name.last name@nottingham.ac.uk}

\renewcommand{\shortauthors}{F. Author et al.}

\begin{abstract}
  As the Internet of Things continues to take hold in the commercial world, the teams designing these new technologies are constantly evolving and turning their hand to uncharted territory. This is especially key within the field of secondary service design as businesses attempt to utilize and find value in the sensor data being produced by connected products. This paper discusses the ways in which a commercial design team use smart thermostat data to prototype an advice-giving chatbot. The team collaborate to produce a chat sequence through careful ordering of data \& reasoning about customer reactions. The paper contributes important insights into design methods being used in practice within the under researched areas of chatbot prototyping and secondary service design.
\end{abstract}


\maketitle

\section{Introduction}

The hyped `vision' for the Internet of Things (IoT) sees large quantities of sensor data being made available from a plethora of ubiquitous connected devices. This data lends itself to the production of secondary services \cite{lee2014service}, also referred to as value-added services \cite{theodoridis2013developing}, which provide auxiliary functionality to devices and services connected to the IoT. The practical reality of designing these services, presents some significant challenges for design teams, as they attempt to make sense of human, social activity from heterogeneous data sets. This paper looks at how a commercial Digital Research and Development (R\&D) team use smart thermostat (ST) data as a design resource to prototype an advice-giving chatbot.

Previous studies in this area have looked at how smart home data can be used, interpreted and accounted for by household members \cite{tolmie2016has} and also how design can support the use of smart home data to provide advice and insights to householders \citep{fischer2017data}. These studies highlight some of the challenges that design teams face when reusing smart home data to design secondary services. This raises the question of how commercial design teams are adapting to these new challenges and incorporating IoT data into their design work. 
\begin{sidebar}
    \includegraphics[width=\marginparwidth]{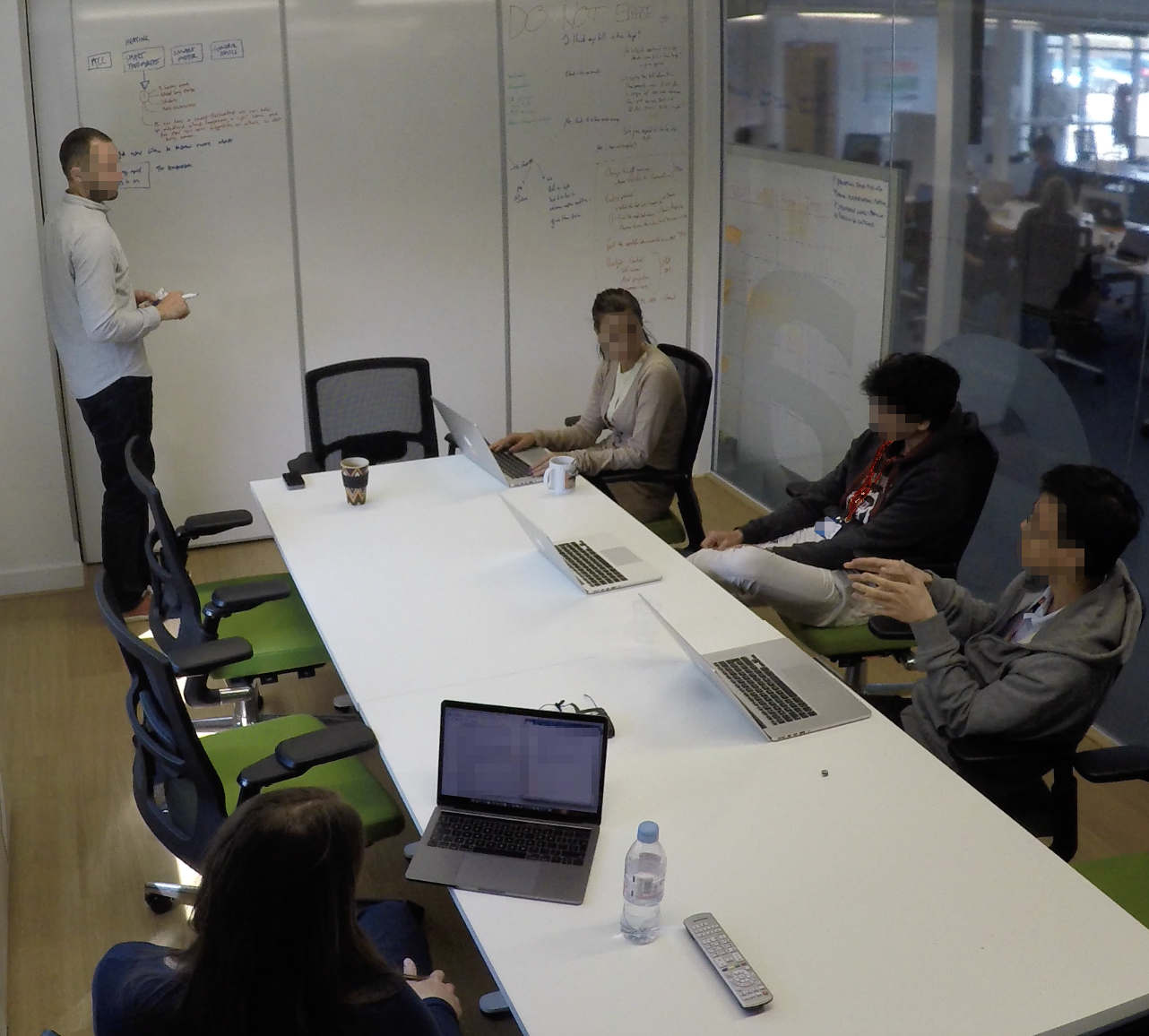}
      \textbf{Sidebar 1: Design meeting in progress. The team are creating a prototype chat sequence on the whiteboard.}
      \vspace{5mm}
      \subsection{Design brief}
To provide information and advice, based on smart thermostat (ST) data, to answer the question 'how can I save money?', in an understandable and customer friendly way.\\
     \textbf{Sidebar 2: The design brief discussed at the start of the meeting}
    \label{bar:sidebar}
    \vspace{5mm}

     \textbf{Transcript 1: The team are discussing the potential of an additional step in the chat sequence to provide context to the user before introducing the boiler on-time data\\}
    1: \textbf{PM:} so by saying things like, your bill will be\\
    2: \hspace{0.6cm}impacted by the time your heating is on \\
    3: \hspace{0.6cm}each month, sorts of things like this.\\
    4: \hspace{0.6cm}Then you're kind of getting the customer\\
    5: \hspace{0.6cm}into the mindset, and \textit{then} we can say \\
    6: \hspace{0.6cm}your boiler was on this much.
  
  \label{bar:sidebar}
\end{sidebar} 
Chatbot interaction design is also an emerging area of HCI research \cite{folstad2017chatbots} where there is a distinct lack of design methods and frameworks that can be utilized by practitioners. There are a number of papers that discuss elements of chatbot design  \cite{abdul2015survey,nica2018chatbot,setiaji2016chatbot}, however, the main focus of these is on the system elements of the design in terms of the natural language processing (NLP) and knowledge retrieval techniques. There is very little discussion about the design of the responder element \cite{abdul2015survey} and no mention of prototyping methods or processes to support commercial design teams. Insight is needed into the design work involved in prototyping chat sequences as a practical accomplishment.

This paper unpacks an R\&D team design meeting, within a large energy company, as they approach the challenge of adding value to smart home data through a prototype chatbot.  

\section{The Study}

The case study reported in this paper forms part of wider work looking at the challenges of design practitioners working with the IoT. The data gathering aspects consisted of ethnographic fieldwork. As an experienced design practitioner, the first author of this paper was embedded as a participant observer throughout the work. The approach to the research data consisted of 3 phases; participation, familiarization with the research data and evaluation of the team's approach to adding value to data. 

To avoid any confusion the following terms have been used: 
\textbf{`Chat'} refers to conversational sequences being designed within the chatbot. 
\textbf{`Data'} refers to the resources the team are working with. 
\textbf{`Research data'} refers to data observed and recorded as part of the research study.

The design meeting discussed in this paper was recorded as part of a larger corpus of data collected during a 2 week sprint. The sprint involved the team focusing on the creation of a prototype to test the technical feasibility of using customer and IoT data to provide advice to customers through a chatbot system. 

It is important to note that the meeting occurred 5 days into the sprint so the team had already familiarized themselves with the systems that they were using and the available data.

There were 3 members of the team invited to the meeting by the Project Manager (PM), which comprised of Liz, Hugo and Neil. Liz had been working on the front end of the chatbot, creating the wording and structure for the chat. Hugo and Neil had been focusing on the data streams and how they could utilize them to provide information and advice. The PM took the lead throughout the meeting, reiterating the design brief (see sidebar 2) at the start and using the whiteboard to record and display elements of the chat sequence as the team worked through it.

The meeting itself focused on the designing of one chat sequence, or \begin{sidebar}

     \textbf{Transcript 2: The team assess the options they would like to offer the user within the chat sequence. Based on this discussion the team restrict the users options at this step and decide to deliver the data in a step by step manner}\\
    1: \textbf{PM:} So we can't offer them the choice\\
    2:\hspace{0.7cm}of looking at how long their heating\\
    3:\hspace{0.7cm}is on, the temperature they've set.\\
    4: \textbf{H:}\hspace{0.2cm} So you, you can definitely do that, but\\
    5:\hspace{0.7cm}the thing is, average set point and\\
    6:\hspace{0.7cm}average outdoor temperature are two \\
    7:\hspace{0.7cm}entities that are used to uh.\\
    8: \textbf{PM:} you're saying that this is a product\\
    9:\hspace{0.7cm}of the other two things?\\
    10: \textbf{H:}\hspace{0.1cm} yes\\
    11: \textbf{L:}\hspace{0.1cm} ((nods))  \\
    12: \textbf{H:}\hspace{0.1cm} I mean, the cause is how long your \\
    13:\hspace{0.6cm}heating is on. I mean, the consequence \\
    14:\hspace{0.6cm}of the average set point and the average \\
    15:\hspace{0.6cm}outdoor temperature is how long your  16:\hspace{0.6cm}heating is on. \\
   17: \textbf{N:}\hspace{0.1cm} and then you answer to the question \\
   18:\hspace{0.6cm}why. Why is the data. What is the data.\\
   19: \textbf{PM:}  ok, so we have to do it differently.\\
   20:\hspace{0.6cm}((rubs out button on whiteboard)) so the\\
   21:\hspace{0.6cm} only thing we can offer then. Well the\\
   22:\hspace{0.6cm}first thing we have to offer them, is  \\
   23:\hspace{0.6cm}information about how long their boiler\\
   24:\hspace{0.6cm}has been running.

  \label{bar:sidebar}
\end{sidebar}'journey', as the team referred to it. The work was purely focused on the conversational elements of the chatbot's user interface (UI). The technical aspects of the chatbot infrastructure had previously been decided upon and were not part of the design work in this meeting. The prototype was an initial iteration designed to get feedback on the UI elements of the project. As such, the design task was to create a chat sequence in a fairly static way. Later iterations were planned to include the NLP \& machine learning elements.
\section{Findings}
One of the key challenges faced by the team was how to make data meaningful so that it added value for the customer.
There were 5 data streams that the team had the potential to use to provide insight to users within the chat sequence. The ST provided boiler on-time (BO) and thermostat set point (SP). Based on the design brief and the initial fixed steps of the journey, it was implied that these core data streams would be included. These were key resources from which the team could draw inferences about activity in the home. Another data stream that was fairly central to this design task was weather data, or more specifically outdoor temperature (OT). This was considered important by the team and its inclusion was discussed on a number of occasions.  

The final 2 potential data streams were Customer and Energy Performance Certificate (EPC) data. These were not included in this journey as they were referred to as having their 'own journeys'. Relationships between journeys was mentioned as something to be reviewed in a future project phase.

\subsection{Assessing relevance}

The team assigned a sequential order to the data in relation to the design brief. This was not mapped out in a formal manner but can be seen throughout the discussions, with an example of this is shown in transcript 1 where BO was introduced as having a large impact on a customers bill (lines 1-3). The assumption here was that if the heating is on for less time then the customer will save money, thus making it directly relevant to the initiating question, 'how can I save money?'. In transcript 2 this sequential order is further cemented. The PM initially (lines 1-3) asks for clarification from Hugo about his justification for blocking a previous suggestion of giving the customer the option to choose which data they would like to look at. Hugo responds to this (lines 4-7) by clarifying that it is possible from a practical perspective. He then presents a justification which highlights why BO is the most relevant to the initiation question (lines 12-16). The PM accepts this justification (line 19) and adopts the view that BO \textit{has to} be presented first (line 22). 

This relationship mapping that the team assigned between the data and the initiation question ultimately provided the structure for the chat sequence. This is in contrast to how interesting the members believe the data is. There are many occasions where data is labelled as 'interesting' or 'more interesting' than another data stream. Transcript 3 shows an argument that states while BO is interesting, SP \& OT are more interesting as they can explain BO (lines 1-3). \begin{sidebar}
         \textbf{Transcript 3: The team are discussing what content to include in the next step of the sequence. The focus is brought back to how the information can be used by the user to meet their goal of saving money. A scenario is used to highlight the users motivation based on the initiation question in the design brief.}\\
         1:\textbf{PM:} So boiler on-time is interesting but the \\
         2:\hspace{0.6cm}reasons for the boiler being on are more \\
         3:\hspace{0.6cm}interesting do you think? Do you agree?\\
         4: \textbf{H:}\hspace{0.1cm} Yeah yeah yeah\\
         5:\textbf{PM:}	Because that's something you can do \\
         6:\hspace{0.6cm}something with or not. Even knowing\\
         7:\hspace{0.6cm} that my boiler was on for 172 hours \\
         8:\hspace{0.6cm} doesn't really help me. It just states the\\
         9:\hspace{0.6cm} fact.\\
         10: \textbf{H:}\hspace{0.1cm} But you have to state it anyway \\
        11:\textbf{PM:}Of course, no I agree that, but knowing\\
        12:\hspace{0.6cm}now. The bit I wanna know, ok fine,\\
        13:\hspace{0.6cm}but what do I do? that's why I'm here,\\
        14:\hspace{0.6cm}I wanna save money so what do I do\\
        15:\hspace{0.6cm}to reduce this? ((taps the BO step on\\
        16:\hspace{0.6cm}the whiteboard))\\
  \label{bar:sidebar}
\end{sidebar}The terms interesting and meaningful are used to refer to data outputs that are insightful or actionable for the customer. This is a type of relevance in itself that links to customer goals and motivations but one that supports the sequential order of the chat journey rather than driving it. By assessing the relevance in these ways the members are able to map the data from the initiation question through to actionable insights in a step by step manner. 

\subsection{Adopting the customer perspective}

One of the key elements of this design task, as specified in the design brief, was that the chat needed to be understandable and customer friendly. By defining the question that initiates the journey the team provide a reasonable sense of the goals and motivations of their user. Designing the prototype in this static manner enables the team to draw on scenarios \& predict reactions related to the customers' goals. This direct approach can be seen throughout the meeting to further assess the relevance, meaning and value of the chat steps in the journey. 

Scenarios are used to reason about how the customer will be utilizing the information presented to them. They can be presented by any member to support an argument. Transcript 4 demonstrates how the PM draws on a scenario to make his doubts about the relevance of the OT data stream accountable to the team. He introduces a scenario by first stating his belief that weather information is only relevant in some circumstances (lines 1-3). A usage scenario is then delivered in the first person perspective giving an example of when the information would be meaningless to a user (lines 6-14). The inclusion of the initiation question (line 8) is used to highlight the customer's goal as he moves between the design brief and the customer perspective to frame his argument. As a result of this discussion the team decide to drop the OT data from this journey. 

Transcript 3 shows an example of another scenario being used to demonstrate how the data meets the assumed goals and expectations of the user (lines 12-15). In this example the scenario is intertwined with other conversational elements. The PM switches between conversation about the data and delivering a scenario. The scenario is made accountable to the other members by the use of the first person perspective, allowing the PM successfully switch his assumed role from designer to user. In this case the purpose of the scenario is not only tied to a specific step in the chat journey but is also used to refocus the team on the design brief and the assumed customer goal, in a more general way, relating to the design task of the journey as a whole. 

In addition to the use of scenarios the team also rely on predictions of customer reactions. These are much shorter than the scenarios and as such, are drawn on more regularly throughout the task. Their purpose is similar, they are used to justify or back up a decision or argument in relation to whether or not it adds value or meaning to the customer. These predicted reactions can be either positive or negative. Transcript 1 shows an example of a positive predicted reaction being presented to support an argument for \begin{sidebar}
     \textbf{Transcript 4: The team are deciding whether or not to include weather data in the chatbot. An example scenario is offered to reason about how relevant and meaningful it would be to a user. As a result of this discussion the team decide not to include the data stream.}\\
    1: \textbf{PM:}so there are some quite specific\\
    2:\hspace{0.6cm} conditions under which weather \\
    3:\hspace{0.6cm} information is relevant to that person's \\
    4:\hspace{0.6cm} visit. \\
    5: \textbf{H:}\hspace{0.1cm} mmhmm\\
    6: \textbf{PM:} because if I've just come on and I go\\
    7:\hspace{0.6cm} through Liz's welcome thing and I see\\
    8:\hspace{0.6cm} how to save money as an option, and I\\
    9:\hspace{0.6cm} click on it, and it's August, knowing\\
    10:\hspace{0.6cm}what the average weather or what the\\
    11:\hspace{0.6cm}average temperature was in the last 30 \\
    12:\hspace{0.6cm}days against the rest of the UK, or last\\
    13:\hspace{0.6cm}30 days against the previous 30 days\\
    14:\hspace{0.6cm}it's kind of meaningless isn't it.
  \label{bar:sidebar}
\end{sidebar}adding a context step before introducing the BO data (lines 4-5). The justification for the additional step is to set the frame of mind of the visiting customer, by making the meaning of the data presented in the following step explicit. In contrast, transcript 5, gives an example of a negative predicted reaction as Hugo suggests presenting the BO data in hours over a 30 day period may cause the customer to \textit{freak out} (line 10). 

\section{Discussion}
The findings demonstrate how the team collaboratively work through the task of creating a prototype chat sequence that adds value to ST data streams. The constraints placed on the task, through the specific design brief and decision to keep the prototype fairly static, frame the task in a way that maintained focus on adding value to the data. The team can be seen to create a sequential order to the data that is used to drive the step by step elements of the design. This approach sees the systematic linking of the data to the design brief as they draw on their knowledge of the data streams. 
The method of creating a step by step mapping of the chat sequence provides some interesting insights into chatbot design work. The sequence mapping seen in this case was performed before any NLP \& ML elements were included, thus giving potential for user testing to be performed on the user UI elements in a somewhat controlled manner. 

The focus on the customer within the design brief appeared to encourage the team to reason about how the customers would be using and reacting to the information that was being presented to them. This ties into some of the challenges discussed by \cite{tolmie2016has, fischer2017data}. Scenarios and predicted reactions were drawn on to justify design ideas and support decisions. At such an early stage of the project, before any user feedback had been received by the team, they used their knowledge about their customers and personal experience to draw on. This technique was central to the design task in respect to how the team were progressing and reviewing their progress in relation to adding value to the data. 
\section{Conclusions and future work}

Design using sensor data to add value through secondary services is a complex task. This study demonstrates how the team find sequential order in the data to drive the structure of the design forward. Value is added through a process of reasoning about how the users will use and react to the information presented to them. Insights are also provided in relation to the work of prototyping of a chatbot UI designed to facilitate future user testing opportunities. 

Future work will look at the wider corpus of data across the whole 2 week sprint, along with additional studies of practice in this area. 

This paper opens up a view into complexities involved in design with IoT data and highlights current methods of commercial chatbot design work. More work is needed to enable HCI to provide support to practitioners in the under \begin{sidebar}

     \textbf{Transcript 5: The team are discussing how to display the boiler on-time data to the customer. They are trying to ensure that the information they present is both meaningful and customer friendly.}\\
        1: \textbf{H:}\hspace{0.1cm} I think it would be days \\
        2: \textbf{L:}\hspace{0.1cm} because if it is for the last 30 days\\
        3:\hspace{0.6cm}hours is going to be pretty big\\
        4: \textbf{PM:}Yeah, well, but days is kind of a weird\\
        5:\hspace{0.6cm}thing to compare boiler on time in, I\\
        6:\hspace{0.6cm}think, hours. You schedule your heating\\
        7:\hspace{0.6cm}in hours rather than days. I don't know \\
        8:\hspace{0.6cm}I think we just need to see how that-\\
        9: \textbf{H:}\hspace{0.1cm} we can display hours but I\\
        10:\hspace{0.6cm}feel the customer will freak out. Because\\
        11:\hspace{0.6cm}sometimes you know it's uh-\\
        12: \textbf{L:}\hspace{0.1cm} massive

  \label{bar:sidebar}
\end{sidebar}researched areas of Chatbot prototyping and secondary service design.

\begin{acks}
    This work is supported by the Engineering and Physical Sciences Research Council [grant number EP/K025848/1]. 
\end{acks}

\bibliography{bibliography}


\begin{thebibliography}{8}


\ifx \showCODEN    \undefined \def \showCODEN     #1{\unskip}     \fi
\ifx \showDOI      \undefined \def \showDOI       #1{#1}\fi
\ifx \showISBNx    \undefined \def \showISBNx     #1{\unskip}     \fi
\ifx \showISBNxiii \undefined \def \showISBNxiii  #1{\unskip}     \fi
\ifx \showISSN     \undefined \def \showISSN      #1{\unskip}     \fi
\ifx \showLCCN     \undefined \def \showLCCN      #1{\unskip}     \fi
\ifx \shownote     \undefined \def \shownote      #1{#1}          \fi
\ifx \showarticletitle \undefined \def \showarticletitle #1{#1}   \fi
\ifx \showURL      \undefined \def \showURL       {\relax}        \fi
\providecommand\bibfield[2]{#2}
\providecommand\bibinfo[2]{#2}
\providecommand\natexlab[1]{#1}
\providecommand\showeprint[2][]{arXiv:#2}

\bibitem[\protect\citeauthoryear{Abdul-Kader and Woods}{Abdul-Kader and
  Woods}{2015}]%
        {abdul2015survey}
\bibfield{author}{\bibinfo{person}{Sameera~A Abdul-Kader} {and}
  \bibinfo{person}{JC Woods}.} \bibinfo{year}{2015}\natexlab{}.
\newblock \showarticletitle{Survey on chatbot design techniques in speech
  conversation systems}.
\newblock \bibinfo{journal}{\emph{International Journal of Advanced Computer
  Science and Applications}} \bibinfo{volume}{6}, \bibinfo{number}{7}
  (\bibinfo{year}{2015}).
\newblock


\bibitem[\protect\citeauthoryear{Fischer, Crabtree, Colley, Rodden, and
  Costanza}{Fischer et~al\mbox{.}}{2017}]%
        {fischer2017data}
\bibfield{author}{\bibinfo{person}{Joel~E Fischer}, \bibinfo{person}{Andy
  Crabtree}, \bibinfo{person}{James~A Colley}, \bibinfo{person}{Tom Rodden},
  {and} \bibinfo{person}{Enrico Costanza}.} \bibinfo{year}{2017}\natexlab{}.
\newblock \showarticletitle{Data work: How energy advisors and clients make IoT
  data accountable}.
\newblock \bibinfo{journal}{\emph{Computer Supported Cooperative Work (CSCW)}}
  \bibinfo{volume}{26}, \bibinfo{number}{4-6} (\bibinfo{year}{2017}),
  \bibinfo{pages}{597--626}.
\newblock


\bibitem[\protect\citeauthoryear{F{\o}lstad and Brandtz{\ae}g}{F{\o}lstad and
  Brandtz{\ae}g}{2017}]%
        {folstad2017chatbots}
\bibfield{author}{\bibinfo{person}{Asbj{\o}rn F{\o}lstad} {and}
  \bibinfo{person}{Petter~Bae Brandtz{\ae}g}.} \bibinfo{year}{2017}\natexlab{}.
\newblock \showarticletitle{Chatbots and the new world of HCI}.
\newblock \bibinfo{journal}{\emph{interactions}} \bibinfo{volume}{24},
  \bibinfo{number}{4} (\bibinfo{year}{2017}), \bibinfo{pages}{38--42}.
\newblock


\bibitem[\protect\citeauthoryear{Lee, Kao, and Yang}{Lee et~al\mbox{.}}{2014}]%
        {lee2014service}
\bibfield{author}{\bibinfo{person}{Jay Lee}, \bibinfo{person}{Hung-An Kao},
  {and} \bibinfo{person}{Shanhu Yang}.} \bibinfo{year}{2014}\natexlab{}.
\newblock \showarticletitle{Service innovation and smart analytics for industry
  4.0 and big data environment}.
\newblock \bibinfo{journal}{\emph{Procedia Cirp}}  \bibinfo{volume}{16}
  (\bibinfo{year}{2014}), \bibinfo{pages}{3--8}.
\newblock


\bibitem[\protect\citeauthoryear{Nica, Tazl, and Wotawa}{Nica
  et~al\mbox{.}}{2018}]%
        {nica2018chatbot}
\bibfield{author}{\bibinfo{person}{Iulia Nica}, \bibinfo{person}{Oliver~A
  Tazl}, {and} \bibinfo{person}{Franz Wotawa}.}
  \bibinfo{year}{2018}\natexlab{}.
\newblock \showarticletitle{Chatbot-based Tourist Recommendations Using
  Model-based Reasoning}. In \bibinfo{booktitle}{\emph{20 th International
  Configuration Workshop}}. \bibinfo{pages}{25}.
\newblock


\bibitem[\protect\citeauthoryear{Setiaji and Wibowo}{Setiaji and
  Wibowo}{2016}]%
        {setiaji2016chatbot}
\bibfield{author}{\bibinfo{person}{Bayu Setiaji} {and}
  \bibinfo{person}{Ferry~Wahyu Wibowo}.} \bibinfo{year}{2016}\natexlab{}.
\newblock \showarticletitle{Chatbot using a knowledge in database}. In
  \bibinfo{booktitle}{\emph{7th International Conference on Intelligent
  Systems, Modelling and Simulation}}.
\newblock


\bibitem[\protect\citeauthoryear{Theodoridis, Mylonas, and
  Chatzigiannakis}{Theodoridis et~al\mbox{.}}{2013}]%
        {theodoridis2013developing}
\bibfield{author}{\bibinfo{person}{Evangelos Theodoridis},
  \bibinfo{person}{Georgios Mylonas}, {and} \bibinfo{person}{Ioannis
  Chatzigiannakis}.} \bibinfo{year}{2013}\natexlab{}.
\newblock \showarticletitle{Developing an iot smart city framework}. In
  \bibinfo{booktitle}{\emph{Information, intelligence, systems and applications
  (iisa), 2013 fourth international conference on}}. IEEE,
  \bibinfo{pages}{1--6}.
\newblock


\bibitem[\protect\citeauthoryear{Tolmie, Crabtree, Rodden, Colley, and
  Luger}{Tolmie et~al\mbox{.}}{2016}]%
        {tolmie2016has}
\bibfield{author}{\bibinfo{person}{Peter Tolmie}, \bibinfo{person}{Andy
  Crabtree}, \bibinfo{person}{Tom Rodden}, \bibinfo{person}{James Colley},
  {and} \bibinfo{person}{Ewa Luger}.} \bibinfo{year}{2016}\natexlab{}.
\newblock \showarticletitle{"This has to be the cats": Personal Data Legibility
  in Networked Sensing Systems}. In \bibinfo{booktitle}{\emph{Proceedings of
  the 19th ACM Conference on Computer-Supported Cooperative Work \& Social
  Computing}}. ACM, \bibinfo{pages}{491--502}.
\newblock


\end{thebibliography}
\bibliographystyle{ACM-Reference-Format}

\end{document}